\documentstyle[12pt]{article}
\begin{document}
\title{Planck Scale to Hubble Scale}
\author{B.G. Sidharth$^*$\\
B.M. Birla Science Centre, Adarsh Nagar, Hyderabad - 500 063 (India)}
\date{}
\maketitle
\footnotetext{$^*$Email:birlasc@hd1.vsnl.net.in; birlard@ap.nic.in}
\begin{abstract}
Within the context of the usual semi classical investigation of Planck
scale Schwarzchild Black Holes, as in Quantum Gravity, and later attempts
at a full Quantum Mechanical description in terms of a Kerr-Newman metric
including the spinorial behaviour, we attempt to present a formulation
that extends from the Planck scale to the Hubble scale. In the process
the so called large number coincidences as also the hitherto inexplicable
relation between the pion mass and the Hubble Constant, pointed out by
Weinberg, turn out to be natural consequences in a consistent description.
\end{abstract}
\section{Introduction}
Recently a description of Fermions was given in terms of the Kerr-Newman
metric, and a background Zero Point Field with a cut off at the Compton
wavelength scale\cite{r1,r2,r3,r4}. Indeed as is well known the Kerr-Newman
metric describes the gravitational and electromagnetic field of an
electron including the anomalous gyro magnetic ratio\cite{r5}. On the
other hand the Zero Point Field with a Compton wavelength cut off not only
avoids divergences, but it leads to the energy of a typical ementary
particle\cite{r2,r6,r7}.\\
It was shown that it is thus possible to think of particle formation from the Zero Point Field,
the particles themselves being Kerr-Newman type Black Holes, but with the
Quantum Mechanical input that the Zitterbewegung effects within the Compton
wavelength remove a naked singularity.\\
This scheme then leads to a cosmology that not only predicts puzzling
empirical relations and is consistent, but infact
predicts an ever expanding universe with a cosmological constant\cite{r1,r8,r9}.
Indeed it is quite remarkable that latest observations by different teams of
observers confirm all this\cite{r10,r11,r12}.\\
At the other end of length scales, it can also be shown that the usual quark
picture can be recovered from the above description\cite{r13}.
\section{The Planck Scale to Hubble Scale}
The above formulation works for Fermions. In Quantum Gravity however, we deal
with Schwarzchild Black Holes, without the all important spin half. This
is a semi classical domain brought out by the fact that for the Planck
mass $m_P \sim 10^{-5}gms$, we have
\begin{equation}
\frac{Gm_P}{c^2} \sim \frac{\hbar}{mc}\label{eA}
\end{equation}
the left side is the classical Schwarzchild radius while the right side
gives the Quantum Mechanical Compton wavelength. From (\ref{eA}) it follows
\begin{equation}
\frac{Gm_P^2}{e^2} \sim 1\label{eB}
\end{equation}
which shows that all the energy at this scale is gravitational.\\
For a typical elementary particle, the pion for example we would have instead
of (\ref{eB}), the well known relation,
\begin{equation}
\frac{Gm^2}{e^2} \sim 10^{-40}\label{eC}
\end{equation}
which shows that at these scales it is electromagnetism that predominates.\\
We will now throw further light on the fact that at the Planck scale it is
gravitation alone that manifests itself. Indeed Rosen\cite{r14} has pointed
out that one could use a Schrodinger equation with a gravitational interaction
to deduce a mini universe, namely the Planck particle. The Schrodinger
equation for a self gravitating particle has also been considered\cite{r15},
from a different point of view. We merely quote the main
results.\\
The energy of such a particle is given by
\begin{equation}
\frac{Gm^2}{L} \sim \frac{2m^5G^2}{\hbar^2}\label{eD}
\end{equation}
where
\begin{equation}
L = \frac{\hbar^2}{2m^3G}\label{eE}
\end{equation}
(\ref{eD}) and (\ref{eE}) bring out the characteristic of the Planck
particles and also the difference with elementary particles, as we will
now see.\\
We first observe that for a Planck mass, (\ref{eD}) gives, self consistently,
$$\mbox{Energy} \quad = m_P c^2,$$
while (\ref{eE}) gives,
$$L = 10^{-33}cms,$$
as required.\\
However, the situation for pions is different: They are parts of the universe
and do not constitute a mini universe. Indeed, if there are $N$ pions
in the universe, then the total gravitational energy is given by, from
(\ref{eD}),
\begin{equation}
\frac{NGm^2}{L}\label{eF}
\end{equation}
As this equals $mc^2$, we get back as can easily be verified, equation (\ref{eC}),
or equivalently we deduce the pion mass!\\
Indeed given the pion mass, one can verify from (\ref{eE}) that $L = 10^{28}cms$
which is the radius of the universe, $R$. Remembering that $R \approx \frac{c}{H}$,
(\ref{eE}) infact gives the empirical and otherwise inexplicable
Weinberg formula\cite{r16},
\begin{equation}
m = (\frac{\hbar^2H}{Gc})^{1/3}\label{e1}
\end{equation}
where $m$ is the pion mass and $H$ is the Hubble Constant.\\
Again gravitation dominates at large scales and the universe itself shows up
as a Black Hole: Not only is $R \sim \frac{GM}{c^2}$, where $M$ is the mass
of the universe, as for the Schwarzchild Black Hole, but also it can be
shown that the age of the universe equals the proper time for travel from
the centre to the edge as in the case of a Black Hole\cite{r9}.
\section{The Stochastic Universe}
The cosmological scheme referred to above, also deduces the various so called
Dirac large number "coincidences", while at the sametime deducing the mysterious
and poorly understood  relation (\ref{e1}). What makes (\ref{e1})
remarkable is the correlation between an elementary particle and a cosmological
constant, called the micro-macro nexus in\cite{r1}.\\
The following suggestion was made to explain all this in reference\cite{r13}:
If we consider the $N$ particles of the universe, as a statistical collection,
then the typical uncertainity length $l$ is given by
\begin{equation}
l = R/\sqrt{N}\label{e2}
\end{equation}
where $R$ is the dimension of the system, in this case the radius of the
universe. As is well known, and this is one of Dirac's large number
coincidences, $l$ in (\ref{e2}) is given by the Compton wavelength of the
typical elementary particle, namely the pion.\\
This brings us right back to the Compton wavelength and the Kerr-Newman
description referred to above which also leads to the various fundamental
interactions as detailed in references\cite{r1,r2,r4} and\cite{r13}.\\
Let us now consider the stochastic picture in a little greater detail. As
is well known, it is possible to obtain the Schrodinger equation from the
theory of Brownian motion\cite{r17,r18,r19,r20}.\\
In this case the motion of particles whose position is given by $x(t)$ is
subject to random stochastic corrections $\Delta x(t)$. The Brownian
character is expressed by the fact that the average of $\Delta x(t)$ in a
time $\Delta t$ vanishes that is,
$$<\Delta x> = 0$$
whereas the random change is given by
$$|\Delta x| = \sqrt{<\Delta x^2>} \approx \nu \sqrt{\Delta t}$$
where the diffusion constant is given by
$$\nu = \frac{\hbar}{m}$$
and is related to the correlation length or mean free path by the
equation
\begin{equation}
\nu = lv\label{e5}
\end{equation}
We can then go on to the usual derivation of the Fokker-Plancke equations.\\
In Nelson's derivation refered to above, the Schrodinger wave function,
as in the De Broglie-Bohm theory (cf.ref.\cite{r2}), is decomposed as
$$\psi = \sqrt{\rho} e^{\imath S/\hbar}$$
The Schrodinger equation itself decomposes into two equations:
$$v = \frac{1}{m} \nabla S$$
$$\frac{\partial S}{\partial t} = -\frac{1}{2m} (\partial S)^2 + V +
\frac{\hbar^2}{2m} \frac{\nabla^2 \sqrt{\rho}}{\sqrt{\rho}}$$
$v$ is identified with the velocity, but there is a new term,
\begin{equation}
V \equiv V_{\mbox{quantum}} = \frac{\hbar^2}{2m}\frac{\nabla^2}{\sqrt{\rho}}
{\sqrt{\rho}}\label{e8}
\end{equation}
It is with this term, which is absent in the classical Hamilton-Jacobi
theory, that Quantum Mechanics diverges from classical theory-- this term
also contains the reduced Planck Constant $\hbar$. As Smolin notes\cite{r19},
"...any attempt to explain Quantum Mechanics as arising from a probabilistic
description of some more fundamental level of dynamics must come down to
an explanation of this term".\\
We would like to point out that all these equations, infact come up
quite naturally in the above Kerr-Newman type Black Holes formulation
(cf.ref.\cite{r2} for details).
To put it simply, with $v = c$, equation (\ref{e5}) above gives for the
correlation length, the Compton wavelength, which is the radius of a
relativistic vortex, or more accurately the Kerr-Newman Black Hole in this
context, while the equation (\ref{e8}) gives the rest energy of the
particle. This could be deduced independantly in the relativistic case
(cf.ref.\cite{r2}).\\
What all this means is that, as pointed out in\cite{r1} the assembly of
$N$ particles leads to a minimum uncertainity length, namely the Compton
wavelength, and similarly uncertainity within the Compton time. In other
words we have to consider minimum space time intervals: One cannot go to
arbitrarily small space time intervals or points. Indeed in Quantum
Mechanics space time points are very debatable concepts, though they have
been used invariably (cf.ref.\cite{r21} and references therein).\\
In the above light, as in Nelson's formulation, we come across what may
be called Stochastic Holism-- each particle is governed by the probability
distribution of the entire assembly. There is a two-way feed back. 
Indeed as pointed out, in references
\cite{r1} and\cite{r21} the supposedly mysterious equation (\ref{e1}) symbolises this
holistic aspect. So also equation (\ref{e2}) and other large number
coincidences including (\ref{eC}) become perfectly meaningful in this
light (cf.refs.\cite{r1,r9}).\\
The underpinning mechanism is an ambient ZPF. At the Compton wavelength
scales, as pointed out above, this shows up as particles. Indeed at the
length scale $l$ the energy density of the background ZPF is given by\cite{r22}
$\sim \frac{\hbar c}{l^4}$ so that the energy within the Compton wavelength
scale $l\quad \mbox{volume}\quad \sim mc^2$, the rest energy of the particle as required.\\
It is again this ZPF which throws up a cosmological constant at large scales\cite{r1,r9,r23}.\\
\section{Discussion}
We now make a few observations:\\
i) It must be remarked that what have hither to been considered to be two
different fluctuations viz., the Quantum fluctuations and the Statistical
fluctuations have really been unified into a single consideration.\\
ii) The $V$ term in equation (\ref{e8}) which characterises Quantum Mechanics
now shows up as the rest energy of particles, the correlation between the
Brownian processes being within the Compton wavelength correlation length
which characterises the particle.\\
iii) Once we have a formulation of particles within the Compton wavelength
as Kerr-Newman type Black Hole, then we not only get a unified picture
of electromagnetism and gravitation\cite{r4} but also, as pointed out in
the introduction the strong interactions and a quark picture emerge (cf.
ref.\cite{r2} and \cite{r13} for details).\\
iv) Incidentally it may be pointed out that from much in the spirit of the
foregoing considerations one could obtain the Planck Constant in terms of
the fluctuations in the particle number $N$ of the universe.\\
The fluctuation in the mass of a typical elementary particle like the pion
due to the fluctuation of the particle number is given by
$$\frac{G\sqrt{N} m^2}{c^2R}$$
So we have
\begin{equation}
(\Delta mc^2) T = \frac{G\sqrt{N}m^2}{R} T = \frac{G\sqrt{N}m^2}{c}\label{e9}
\end{equation}
as $cT = R$. It can be easily seen that the right side of (\ref{e9}) equals $\hbar$!
That is we have
\begin{equation}
\hbar \approx \frac{G\sqrt{N}m^2}{c},\label{e10}
\end{equation}
The equation (\ref{e10}) expresses the Planck constant in terms of non
Quantum Mechanical quantities.\\
v) The considerations of Section 2 indicate why the Planck particles at the
lowest scale and pions at the elementary particle scale are fundamental.
Thus equation (\ref{eD}) immediately leads to the Planck mass and the Planck
length for a single particle universe, whereas equation (\ref{eF}) for a
$N$ particle universe leads to the pion mass and the radius of the universe.
This can also be seen in the light of the background ZPF. For a single
particle universe, equating the gravitational energy, given by the left side
of (\ref{eD}) with the ZPF energy in a volume $\sim L^3$, where now $L$
is the Planck length, viz., $\hbar c/L$, we get back (\ref{eA}) or
equivalently the Planck mass.\\
However equating the corresponding energy for an elementary particle given
by (\ref{eD}) with a similar ZPF energy and using (\ref{e2}), we recover
this time the mass of the pion, which thus shows up as a natural choice
for a typical elementary particle.\\
vi)In the light of the comments in Section 3 the Quantum non locality is no
longer mysterious\cite{r21,r24}.


\begin{thebibliography}{99}
\bibitem {r1} Sidharth, B.G., (1998) Int.J. of Mod.Phys.A
13(15), pp599ff.
\bibitem {r2} Sidharth, B.G., (1997), Ind.J. Pure \& Appd.Phys., \underline{35} (7),
456-471.
\bibitem {r3} B.G. Sidharth, (1997)  Non Linear World, 2 (4), pp157ff.
\bibitem {r4} Sidharth, B.G., (1998) Gravitation \& Cosmology, Vol.4, No.2, pp158ff.
\bibitem {r5} Misner, C.W., Thorne, K.S., and Wheeler, J.A., (1973), Gravitation,
Freeman (San Francisco).
\bibitem {r6} Feynman, R.P., and Hibbs, A.R., (1965) "Quantum Mechanics and Path
Integrals", McGraw Hill, New York,p245ff.
\bibitem {r7} L.De La Pena., (1983) "Stochastic Processes applied to Physics...", Ed.,
B. Gomez, World Scientific, Singapore,p435ff.
\bibitem {r8} Sidharth, B.G., "Quantum Mechanical Black Holes: An Alternative
Perspective", (1998) in "Frontiers of Quantum Physics", Eds., Lim, S.C., et al, Springer-
Verlag, Singapore.
\bibitem {r9} Sidharth, B.G., (1998) International Journal of Theoretical
Physics, 37 (4), 1307-1312.
\bibitem {r10} Branch, D., (1998), Nature \underline{391}, 23-24. (cf. also Science, Feb.27, 1998).
\bibitem {r11} Perlmutter, S., et.al. (1998), Nature \underline{391}, 51-54.
\bibitem {r12} Coles, Peter and Ellis George, F.R., (1997) "Is the Universe Open
or Closed?", Cambridge University Press, Cambridge.
\bibitem {r13} Sidharth, B.G., (1998) in "Frontiers of Fundamental Physics",
Universities Press, Hyderabad (In Press).
\bibitem {r14}Rosen, N., (1993) International Journal of Theoretical Physics,
32 (8), 1435-1440.
\bibitem {r15} Sidharth, B.G., and Popova, A.D., (1996), Differential Equations
and Dynamical Systems, 4 (3/4), 431-440.
\bibitem {r16} Weinberg, S., (1972), Gravitation and Cosmology, Wiley, New York.
\bibitem {r17} Nelson, E.,(1966) Phys.Rev. 150, 1079ff.
\bibitem {r18} Nelson, E., (1979) in "Lecture Notes in Physics", Vol.100, Springer,
Berlin.
\bibitem {r19} Smolin, L., (1986) in "Quantum Concepts in space and time", Eds.,
R. Penrose, R., and  Isham, C.J., Clarendon Press, Oxford.
\bibitem {r20} Landau, L.J., (1988) in "Statistical Mechanics", Ed., A.I.
Solomon, World Scientific, Singapore.
\bibitem {r21} Sidharth, B.G., (1998) "A New Approach to Locality and Causality", Vigier
Symposium, Canada. (Also xxx.lanl.gov. quant-ph 9805008).
\bibitem {r22} Wheeler, J.A., (1968) "Superspace and the Nature of Quantum
Geometrodynamics", Battelles Rencontres, Lectures, Eds., B.S. De Witt and
J.A. Wheeler, Benjamin, New York.
\bibitem {r23} Zel'dovich, Ya. B., (1968), Sov.Phys.Usp. 11(3), pp381ff.
\bibitem {r24} Gaeta, G., (1993) Phys.Lett.A. 175, 267-268.
\end{thebibliography}
\end{document}